# A REVIEW ON APPLICATIONS AND CHALLENGES IN INTERNET OF THINGS

Subhash Bhagavan Kommina , Kiran Kumar Pulamolu[y],P Kumar Umesh[z], S Sravya[z]
Assistant Professor
Computer Science & Engineering
Sasi Institute of Technology & Engineering
Andhra Pradesh,India
Email: subhash@sasi.ac.in

[y]Associate Professor
Computer Science & Engineering
Sasi Institute of Technology & Engineering
Andhra Pradesh,India
Email: kiran@sasi.ac.in

[z]Computer Science & Engineering Sasi
Institute of Technology & Engineering
Andhra Pradesh,India

[x]Computer Science & Engineering
Sasi Institute of Technology & Engineering
Andhra Pradesh,India

*Abstract*—World has been introduced with a new way of living by inventing this advanced technology Internet of Things (IOT). By using this technology all the real-world objects can be made to interconnect and communicate with each other which can be done by diagnosing, sensing and networking over the internet to fulfill some objectives. Due to advancements that have taken place in sensors, wireless communication systems and cloud computing has brought steps to invent this innovative technology. This paper deals with the elements, applications and the challenges faced by Internet of Things. This paper mainly concentrates on the current scenario of IoT.

Keywords: Internet of Things (IoT), Sensors, Cloud Computing.

## I. INTRODUCTION

Internet is a buzz word in today's world. So, basically one can say internet protocol is a technology which was developed in the 1950's and from then it is developing day by day to an extent where everyone can have their required information by just typing their queries without having any flaws [4]. So then IOT(Internet Of Things) is one of the modest and superior technology which includes other technologies to fulfill its needs. By using IOT there will be an exchange of data between the things and with the sensors and the data gained by exchange will be uploaded to cloud using the internet. For this reason this technology is basically termed as IOT. Everything will have communication with the other things, when the things are equipped with the sensors. One can choose a particular type of sensors for particular type of things. The different kinds of sensors are level sensors, smoke sensors, proximity sensors, pressure sensors and others.

The following are some of the technologies which are used in the IOT

1) RFID Technology
2) ZIG-BEE Technology
3) BARCODE Technology
4) NFC, BLUETOOTH, WIFI
5) SENSORS

The above elements are explained in Section 3. So by using the above technologies, one can make the things to communicate with each other. And after communicating cloud is used to store the information as it's a virtual memory and can be accessed throughout the entire world.

## II. IoT ARCHITECTURE

The architecture of an IOT is classified in four different stages. The first stage, i.e., stage 1 consist wireless networked things like actuators and wireless sensors. The next stage consists of aggregation of data which is obtained from the sensors and as the data will be in the form of analog way, it will be converted into the digital way i.e., we can call it as analog to digital conversion. The next stage i.e., stage 3 ,here preprocessing of the data will be done and later on after the completion of preprocessing, the data will be stored in the cloud or it will be moved to the data centre [6]. In the stage 4, the data will be stored on traditional backend data centre systems after analyzing and managing of data Fig 1 [26].

### A. Sensors and Actuators

Sensors can also be termed as transducers. A device which converts one form of energy to another form is called transduc-ers. transducer. The physical phenomenon will be converted in the form of an electric impulse. This electric impulse will then be interpreted which is used for the determination of reading. Another kind of transducers that are available in numerous IOT frameworks is an actuator. An actuator works in the reverse manner of a sensor. It takes the electrical information and transforms it into physical activity. For example, an electric engine, a hydraulic framework is the whole unique sorts of actuators fig 2.

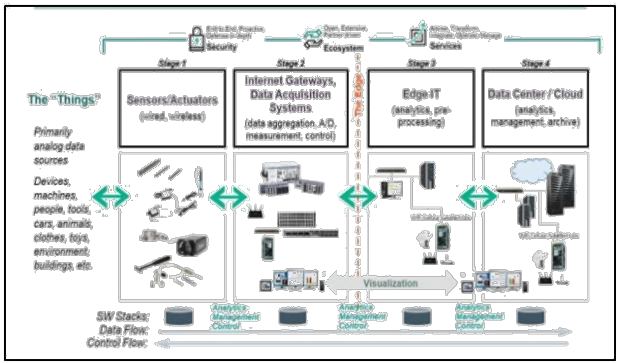

Fig. 1. IoT Architecture

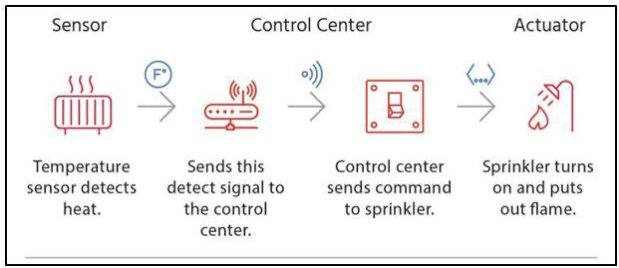

Fig. 2. Sensor to Actuator flow

In IOT frameworks, a sensor may gather data and then routed to a control focus where a choice is made and a required command is sent back to an actuator on basis of sensed input.

### B. Data Aggregation

The information from the sensors begins in the form of analog. That information should be collected and changed over into computerized streams for processing of further downstream. Data acquisition systems (DAS) play out these information accumulation and transformation capacities. The DAS associated with the sensor organize, and plays out the analog to-computerized transformation. The Internet gateway gets the aggregated and digitized information and courses it over Wi-Fi, wired LANs, or the Internet, to Stage 3 frame-works for additionally preparing

### C. Edge IT

When IOT information has been digitized and totaled or aggregated, it is prepared to cross into its domain. But it may, the information may require additionally handling before it enters the server farm. This is the place edge IT frameworks, which perform more investigation, become possibly the most important factor [18]. Edge IT handling frameworks might be situated in remote workplaces or other edge areas, however for the most part these sit in the office or area where the sensors reside nearer to the sensors, for example, in a wiring wardrobe.

Machine learning is at the edge to filter for oddities that recognize approaching upkeep issues that require prompt consideration. At that point everyone could utilize representation innovation to exhibit the data utilizing in straightforward dashboards, maps, or charts. Profoundly incorporated processes frameworks, for example, hyper-met foundation, are in a perfect world suited to these errands since they're moderately quick, and simple to convey and oversee remotely.

### D. Cloud Computing

The Internet of Things (IOT) includes the web associated gadgets that are used to play out the procedures and administrations that help one's lifestyle. Another part set to enable IOT to succeed is cloud computing, which goes about as a kind of front end. Distributed computing is an inexorably prevalent administration that offers a few favorable circumstances to IOT, and it's in view of the idea of enabling clients to perform typical figuring undertakings utilizing administrations conveyed totally finished the internet fig 3. A laborer may need to complete a noteworthy venture that must be submitted to a chief, however maybe they experience issues with memory or space imperatives on their registering gadget. Memory and space requirements can be limited if an application is rather facilitated on the web [10].

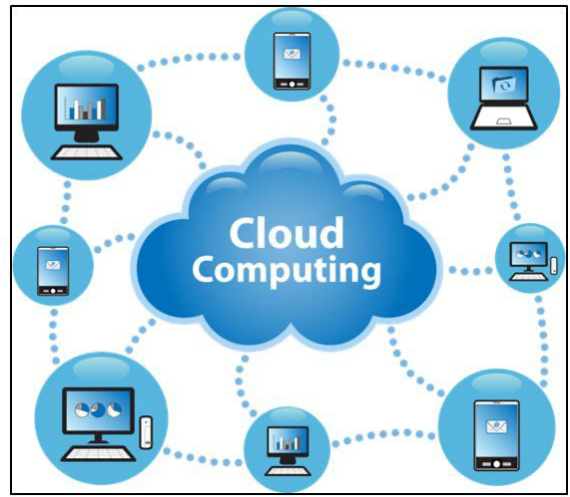

Fig. 3. Cloud Computing in IoT

The specialist can utilize a distributed computing administration to complete their work in light of the fact that the information is overseen remotely by a server. Another case: you have an issue with your cell phone and you have to reformat it or reinstall the working framework. You can utilize

Google Photos to transfer your photographs to web based capacity. After the reformat or reinstall, you would then be able to either move the photographs back to you gadget or you can see the photographs on your gadget from the web when you need.

Initially, the cloud computing of IoT is an on-request self administration, which means it's there when you require it[6]. cloud computing is an electronic administration that can be gotten to with no exceptional help or authorization from other individuals; be that as it may, you require at least a type of web access.

Second, the cloud computing of IOT includes expansive system get to, which means it offers a few availability choices. Cloud computing assets can be gotten to through a wide assortment of web associated gadgets, for example, tablets, cell phones and portable PCs. This level of accommodation implies clients can get to those assets in a wide assortment of behavior, even from more seasoned gadgets. Once more, however, this underscores the requirement for arrange get to points.

Third, cloud computing considers asset pooling, which means data can be imparted to the individuals who know where and how (have consent) to get to the asset, whenever and anyplace. This loans to more extensive joint effort or nearer associations with different clients. From an IOT point of view, similarly as we can without much of a stretch allocate an IP deliver to each "thing" on the planet, we can share the "address" of the cloud-based ensured and put away data with others and pool resources.

Fourth, cloud computing highlights fast flexibility, which means clients can promptly scale the support of their needs. You can without much of a stretch and rapidly alter your product setup, include or evacuate clients, increment storage room, and so on. This trademark will additionally enable IOT by giving versatile figuring force, stockpiling and networking.

At long last, the cloud computing of IOT is a deliber-ate administration, which means you get what you pay for. Suppliers can without much of a stretch measures utilization insights, for example, stockpiling, handling, data transmission and dynamic client accounts inside your cloud occasion. This compensation per utilize (PPU) demonstrate implies your costs scale with your utilization. In IOT terms, it's practically identical to the regularly developing system of physical articles that component an IP address for web availability, and the correspondence that happens between these items and other web empowered gadgets and frameworks; simply like your cloud benefit, the administration rates for that IOT foundation may likewise scale with use.

## III. IoT Elements

Some of the enabling technologies used in this IoT are RFID, ZIGBEE, BARCODE, NFC, BLUETOOTH, WIFI and SENSORS. Among these technologies RFID plays a vital role for the construction of Internet of Things.

### A. Radio Frequency IdentificationRFID

Radio Frequency Identification (RFID) was firstly used in Brittan during the period of 2nd world war . At that time this is used to identify even if the person is spy or ally in 1948. Further in the year of 1999 this technology was reused at Auto-ID centre in MIT. Radio Frequency Identification (RFID) is a technology used to pinpoint an obstacle or a person using radio waves which is done wirelessly. So by using this RFID technology in IOT objects can be easily detected. Tags are attached to the objects to identify using RFID. The major components of this RFID technology are tags, readers, antenna and software. RFID has a wide range of applications in areas such as digital healthcare, Inter vehicular communication, military apps and many more. Challenges of this RFID technology are collision problem, security and privacy concerns[8].

### B. ZIGBEE

ZIGBEE is one of the protocols with some advanced features of wireless sensor networks which is aimed at remote control and sensor applications. ZIGBEE technology was created by ZigBee Alliance which is endowed in the year 2001. Reliability, low cost, low data rate, relatively short transmission ranges are few characteristics of this technology. Applications of this technology are home automation, digital agriculture and power systems. Some of the challenges of this are resource constraint, limited range.

### C. Barcode

Barcode is a contrast approach of encoding letters and num-bers which includes order of bars and spaces with variation in width. In the year 1974, the initial product with barcode was scanned at a check-out counter. These barcodes are machine-readable. Barcode scanners are used to identify a specific object based on the barcode labeled on the item or product. Retail supply chain, shipping and delivery are few applications of barcode technology. Poor security may leads to devastating consequences.

### D. Bluetooth

Bluetooth is a short-range radio wireless technology which is simply used to transfer data between the devices with limited distance. A Personal Area Network can be created by using Bluetooth. This Bluetooth is a wireless surrogate to RS-232 data cables. This is handled by Bluetooth Special interest Group (SIG), which comprises of more than 30,000 member companies in the fields of telecommunication, computing, networking and consumer electronics.

### E. Near Field Communication

Near Field Communication (NFC) one of the wireless communication technologies which is limited to a very short ranges [14]. In 2002, some of the companies named Sony and NXP semiconductors popularized this NFC technology. After 2 years in 2004, many major telecommunication companies associated together for development of this technology. By

2006, several devices such as NFC tags, smart postures and smart tags were imported. The term near communication itself indicates that the communication can be established by positioning the devices close to each other. So simply by touching the two devices one can establish a wireless connectivity without any explicit interaction. NFC makes it easier to create smart environments. Challenges are difficult to ensure the connectivity between the obstacles and security issues [5].

F. Wireless Fidelity

Wireless Fidelity (Wi-Fi) is one of the networking technologies that allows computers, smart phones, tablets and some other devices to communicate with other devices over a wireless signal. In the year 1997, a committee organized by Vic Hayes actualized the 802.11 standard. So, Vic Hayes is the father of wireless fidelity. As the devices are connected wirelessly this will provide applications like smart health care, home automation and smart retail[30].

G. SENSORS

IOT working is completely based on the sensors. Sensors are the devices which takes the input from the surrounding environment and detects it and produce a human-readable output or transmit it over a network for further processing[4] [3]. There are different types of sensors which are based on the Micro Electro Mechanical Systems (MEMS). Some of them are

1) Pressure Sensor: It is a device which takes the pressure as input and later on it converts the taken input which is in the form of pressure into an analogue electric signal. Based upon the pressure applied its magnitude depends. So pressure sensor can also be termed as the pressure transducers as these converts pressure into the electrical energy.

These sensors will be very useful in the fields like aviation, touch screen devices, automobiles, bio medical measurements and many more.

a) Aviation: In this aviation pressure sensors are needed for the airplanes for maintaining the balance between atmospheric pressure and the control systems of the airplanes. Because of this circuitry and various internal components are been protected.

b) Automobile Industry: In this category, pressure sensors are used for monitoring the oil time to time and regulate the power which is used for achieving required speeds when the strong pressure is applied on the accelerator.

c) Touch Screen Devices: All the electronic devices which have touch screen displays will come with the pressure sensors. The electric signal is generated whenever the pressure is applied on the touch screens by using either the stylus or by the fingers which informs to the processor.

d) Bio Medical Measurements: Falling to this category, pressure sensors are used for optimizing the blood pressure which is obtained from the devices like digital blood pressure and ventilators according to the health condition of the patient.

2) Temperature Sensor: The temperature sensor monitors the neighboring thermal reading originated from any obstacle or system which makes one to be able to feel the substantial variation to the temperature generating analogue or digital output. Example for temperature sensor is the ADT7320.Theis temperature sensor which is of digital ranges from -40degree centigrade to +150 degrees centigrade. It is directly combined with micro controller which is located in the IOT device. Need of temperature sensor: The major purpose of this temperature sensor is to detect temperature or heat. The temperature sensor is used to calculate or measure the heat generated by a heat body as well as cold body. It enables us to sense the temperature changed by an object produces a digital or else a analogue outcomes. Based on the application temperature sensors are classified into two types.

a) Contact Temperature Sensor:

This contact temperature sensor which is used to sense the temperature generated by solids, liquids and gases.

b) Non Contact Temperature Sensor: This non contact temperature sensor which is used to sense the temperature or heat generated by the liquids and gases which emits the energy which can be propagated from a body which is in the form of infrared radiation from the sun.

These two temperature sensors are sub classified into 3 forms are given below that are electro mechanical, resistive and electronic.

3) Proximity Sensor: These sensors are used to notice the presence of the obstacles without having any collision. This proximity sensor radiates an electromagnetic field and stays for manipulation in the return signals.

Need of proximity sensors

Ease of detection: By using this proximity sensor the objects can be easily identified without having any physical interaction.

Long lasting life-time: These sensors are having the nature of more reliability and vast utility time as there is no striking between the sensed and sensor objects.

The three major challenges that may be faced by using the sensors are high power consumption, inter operability and security.

IV. IoT APPLICATIONS

Internet of things which is one of the emerging technologies has many real world applications[13]. These applications can be sub-divided based on a lot of factors termed to be category of network availability, area covered, reiteration, user involve-ment [2].

A few applications territories that are to be stated are:
1) Wearable's
2) Home automation
3) Smart cities
4) Smart manufacturing.

A. Wearable Technology

Wearable technology is publicized as one of the tremendous applications of this Internet of Things. Some of the wearable devices imported are smart watch, glasses. These wearable have a meteoric requirement in markets throughout the Globe. Many organizations like GOOGLE, SAMSUNG have entrusted huge budgets to bring out these devices. The functionality of these devices is based on the sensors and software installed. The software are helpful for collecting the data related to end user. The information gathered is refined to get awareness about the user[27].

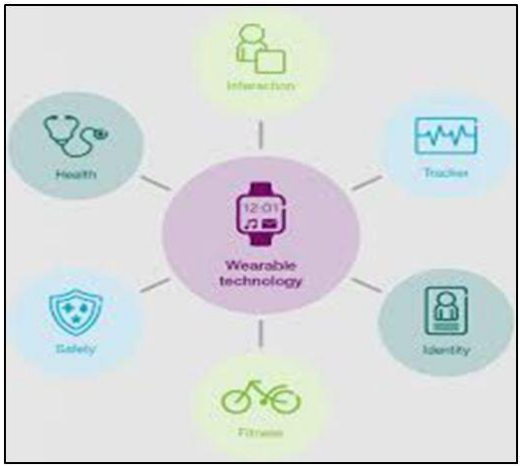

Fig. 4. Wearable Technologies

These wearable are related to fitness, health and entertainment. By using these wearable developed with the help of IOT helps to improve healthcare system [7]. With the help of these devices one can monitor human body constantly. All the information relevant to health conditions such as diagnosis, therapy, medication, weight management, amount of water required and even some routine activities are gathered, managed and maintained efficiently. The heartbeat rate of a patient can be monitored constantly by these wearable and later on this can be sent to doctor[7]. Entire process can be fulfilled by using any computing device with both the end user and doctor, internet access and IOT based healthcare utility. The expansion of the usage of mobile internet service has become a great part of the development of the IOT-powered in-home healthcare (IHH) services fig 4.

### B. Home Automation

Home automation is a method of regulating home appli-ances beyond one's control. The electric and electronic gadgets in the home such as fans, LED's, lights, washing machines, refrigerators and many more can be regulated by applying variety of alarms, kitchen timer control techniques [16]. Smart home has turned into a progressive step of achievement. It is to be anticipated that the smart homes will be as basic as smart phones. There are different methods to manage home appliances such as IOT based home automation over the cloud, home automation under WI-FI through android applications accessed through any smart phone, Arduino based home automation, home automation using digital control.

Among these home automated techniques the most contem-porary one is home appliances remotely over the cloud. In this system Wi-Fi is a wireless networking technology serviced for the interchanging of information in middle of two or more appliances. Home automation means putting on the light naturally without having any human intervention when it gets gloomy. One can easily put off/on the electronic gadgets across any where simply by using smart phones[24]. The sensors will read the current weather, surrounding temperature and humidity levels. Accordingly, to those readings the accurate temperature is maintained across the home. And if a person enters into the house then the air conditioner will be turned on automatically to decrease the temperature of the house. Similarly, there are many such kind of techniques that are useful for home automation. This home automation will be useful save time, energy and wealth[19].

### C. Smart Cities

Smart city is one among the some of the powerful applica-tions of IOT which has created a bit of eagerness among the world's population[33]. Some of the examples of these smart cities are automated transportation, urban security, environ-mental monitoring and smarter energy management systems fig 5 .

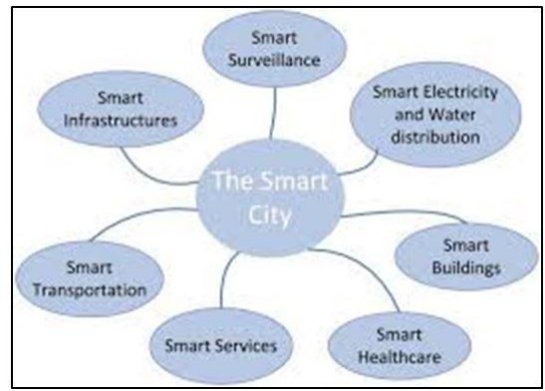

Fig. 5. Smart City Services using IoT

In cities, the main contributor for noise pollution, degra-dation of air quality and radiation of green house gasses is the traffic in urban areas[22]. The automated transportation innovated with the help of Internet of Things may be a bit helpful to reduce the pollution up to some extent. Automated transportation can be made successful by self-driving vehicles. These vehicles can monitor the enclosing area on the roadway and controls the steering, accelerator and breaking system without having any interaction with human beings [29]. These smart cities also include maintaining the traffic levels across different areas and guiding users with the help of mobile applications to choose a path with less denser traffic[20].

The main objective of smart energy management system is to invent a systematic manner which is helpful keep super-vising of every appliance and getting details of all appliance power utilization framework. Later on, the acquired information regarding specific device will be passed through a gateway where an intelligent set of instructions will be maintained to keep track of the parameters as per user requirements. These parameters of distinct individuals can be noticed with the help of an android smartphone.

Simply, by positioning the sensors and using web applica-tions people in urban areas can asset a convenient parking position with in the city. Sensors are also useful to identify installation problems in the electricity system.

Environmental monitoring is also a part of the smart cities which encourages for the protection of the environment by observing the quality of air or water, soil or atmospheric conditions. Monitoring of these factors will be helpful to detect some natural calamities such as tsunami, earthquake. Earlier detection of these calamities will be useful for emergency services to provide valid service [31].

### D. Smart Manufacturing

The Internet of Things has been leading to many unexpected manipulations in business models, integrating the productivity and robotizing the processes in large number of industries. This manufacturing sector has a great effect by this tech-nological revolution [28]. Many manufacturing sectors such as automotive, chemical, electronics, durable goods and many more have invested huge amounts in IOT devices fig 6.

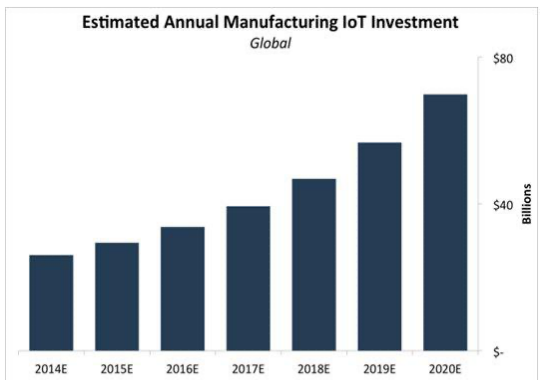

Fig. 6. Smart Manufacturing

Simply, smart manufacturing is termed as utilizing the IOT devices for advancing the adaptability and productivity of manufacturing procedures. This includes installing sensors to the current manufacturing setup. The advanced manufacturing equipment includes pre-installed sensors. Smart manufacturing includes all the fields of business, blurring the boundaries among plant operations, supply chain, product design and demand management[9].

### E. Industrial Internet

Industrial Internet is a latest hum in the industrial zone. This industrial internet is a dub of Industrial Internet of Things (IIOT). This includes the task of empowering industrial engi-neering with sensors, software and big data analytics which are useful for creating sparkling machines[2] [12]. The main aim of introducing this IIOT is to invent smart machines that are more systematic and rational when compared to individuals in the way of approaching the data. This data can empower the organizations to identify the difficulties and inefficiencies instantly which is helpful to save property and time fig 7.

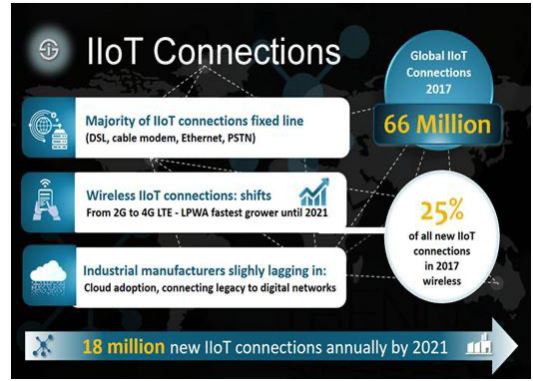

Fig. 7. IIOT Connections

This IIOT is helpful for monitoring and interacting with any device without having any physical interaction which means that the equipments within the industry can be accessed from anywhere across the world by simply having the internet connectivity[4]. For example, Functional manager can monitor specialized equipment functionality from any place, at any time[25].

### F. IOT in Agriculture

As there is an endless raise in world's population, there is a huge demand for food supply. Many advanced techniques have been introduced by the governments which are advantageous for the agriculturists to double the food productivity [1]. Smart farming is one among the rapid expanding sector in IOT. Now-a-days peasants are utilizing the data for getting integration over the investment fig 8.

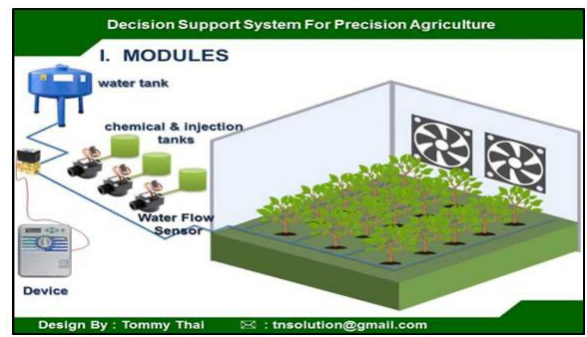

Fig. 8. IoT in Agriculture

This smart agriculture provides the opportunity for the farmers to supervise the soil moisture and nutrients, managing the water utility for maintaining the excellent growth of the crop and identifying the required fertilizer depending on nature of the ground. Many innovative sensors are designed to make a note of soil temperature, humidity and air conditions which are advantageous for identifying best suite crop[23].

## V. CHALLENGES IN IoT

Internet of Things which is one of the emerging technologies has been still in the stage of inception[2]. This emerging technology has been experiencing a lot of challenges. These challenges must be conquered to get the best results of this technology. Some of the challenges are security, connectivity, compatibility and longevity[17].

### A. Security

Widespread of this new technology and its maintenance will correspondingly depend on the security associated with the data[11]. A huge number of things are in touch with each other using web and still a lot of devices are ready for establishing the connection in coming years. The task of the security providers will become complex as large number of devices are connected using internet.

The major cause for security issue is lack of standards for sharing and protecting the information. The provoke of smart fridges, drug infusion pumps, baby monitors and even the radio system in the car are representing the security horror which may takes places in near future of IOT. Large number of new hubs is being added to networks and web may cause nasty actors with numerous attacks which may lead to security holes[32].

Once if there is a solution for the challenges faced due to the security issues then this emerging technology Internet of Things will be more implanted in the lives of human beings[15]. Some of the solutions to ensure the secure con-nected products are:

> At point when a gadget is firstly started up, a cryptographically created digital autograph certify the authen-ticity and integrity of the software on the gadget which indicates that only the software that has been certified to be accessed on that device and entered by the entity that approved it will be stacked.
> When a gadget is connected to the network, it should validate itself prior to acquire or transference of data.
> The gadget must have a firewall check-up scope to restrict traffic that is intended to halt t the gadget in such a manner that makes optimal use of the limited computational resources.

Yet, there are many other organizations working on to design the platforms that make huge networks of IOT gadgets to detect and authenticate each other in order to contribute higher security[2]. Research is also being carried out to improve the IOT security with the help of device and Smartphone linking.

### B. Connectivity

Connectivity is one among the huge challenges to be faced which tends to connect large count of gadgets. The connectivity between the nodes is established by depending on centralized, client/server model. These architectures are suitable for those devices which are a part of the IOT currently. But it is estimated that in upcoming days nearly hundreds of billions of gadgets will unite to network which will turn into a blockage. The capability of the present cloud servers is too less so it may lead to failure of services if it is forced to handle huge amount of data.

The fate of IOT will specifically require to rely upon decentralizing IOT systems. Some portion of it can wind up noticeably conceivable by moving some of the tasks to the edge, such as utilizing fog computing methods where smart gadgets such as IOT hubs undertakes the responsibility of mission-critical operations and cloud servers collect on information gathering and analytical obligations.

Few more solutions includes implementing peer-to-peer communications, in which devices can track and authenticate the other gadgets precisely and interchange data without interacting with any other third party. Mesh networks can be established without having any single breakdown.

### C. Compatibility and Longevity

There is a vast growth of IOT in various directions, with variety of technologies such as RFID, Bluetooth, Wi-Fi and many more which are competing with each other to become definitive[2]. This will lead to many problems and desire the stationing of additional hardware and software at the time of establishing connectivity among the devices. Some other compatibility problems may takes place due to non-unified cloud services, no proper standardized machine to machine protocols and distinction in firmware and operating systems among the IOT devices.

Some of these advances will inevitably end up plainly old in the following couple of years, adequately rendering the gadgets actualizing them useless. This is particularly vital, since as opposed to non specific processing gadgets which have a life expectancy of a couple of years, IOT machines, (for example, shrewd refrigerators or TVs) have a tendency to stay in benefit for any longer, and ought to have the capacity to work regardless of the possibility that their producer leaves benefit[21].

## VI. CONCLUSION

In light of this investigations and report of the Internet Of Things (IOT), it can be stated that this division is simply in the initial steps of improvement and have a tons of potential advancement. This report states the current status of IOT research by criticizing the present trends and challenges. This IOT expands on current technologies termed to ne RFID and WSN along with a few principles and protocols to handle machine-to-machine interaction. The eventual fate of IOT is basically boundless due to the advances in automation and purchaser's wish to coordinate gadgets such as smart phones with household appliances.